# Reducing dislocation defect levels via sub-melt nanosecond pulsed-laser induced densification of diamond

Adam H. Khan[1], Tae Sung Kim[1], Gabe Guss[2], Ted A. Laurence[2], Sonny S. Ly[2], Thejaswi U. Tumkur[2] and Afaq H. Piracha*[1]

[1]Diamond Quanta, 319 North Bernardo Avenue, Mountain View, California, USA

[2]Lawrence Livermore National Laboratory, 7000 East Avenue, Livermore, California, USA

*Corresponding Author

E-mail: afaq@diamondquanta.com (A. H. Piracha)

**Abstract**

Dislocations and polishing-induced defect networks in synthetic diamond crystal introduce local strain fields and broaden Raman features, limiting performance in optical, thermal, and electronic applications. Laser annealing is emerging as a promising way to repair surface or crystal defects in diamond without entering the melt regime, yet the surface densification, defect depletion, and functional property recovery has not been well quantified. In this work, we demonstrate that sub-melt nanosecond pulsed-laser annealing (PLA) can directly reduce dislocation-associated strain by densifying and reorganizing the damaged near-surface region of single-crystal Chemical Vapor Deposition (CVD) diamond. Single- and two-pulse PLA were applied, and structural evolution was quantified using co-registered ISO 25178 white-light interferometry, depth-resolved Raman spectroscopy, and cross-sectional STEM with geometric phase analysis (GPA). Across a 5 × 6 grid (n = 30), responsive regions show large reductions in local slope (Sdq 45-65%), developed area (Sdr 60-90%), height spread (Sp, Sz 30-65%), void volume (Vv 57-60%), and roughness amplitude (Sa, Sq 48-57%), consistent with densification of ~4-6.5 nm. Raman profiling reveals narrowing of the diamond line and improved spectral uniformity to depths of ~2-3 μm, indicating relaxation of dislocation-mediated strain beyond the compaction layer. STEM-GPA strain maps confirm this behavior, showing smoother strain fields, suppressed hotspots, and redistribution of localized strain concentrations following PLA. These results establish that sub-melt PLA provides a direct pathway to reduce dislocation-defect levels in diamond by compacting surface-connected free volume and reducing the strain fields that stabilize or nucleate dislocations. The method offers a scalable approach for upgrading industrial-grade diamond (ranging homoepitaxial and heteroepitaxial CVD single crystal to polycrystalline diamond) toward low-defect, device-ready surfaces, with implications for high-power electronics, photonics, and quantum-grade substrates and wafers.

## 1. Introduction

Diamond's extreme thermal conductivity, wide bandgap, high breakdown field, stiffness, and radiation hardness make it a compelling platform for power microelectronics, photonics, and quantum technologies [1]. Device-scale performance is often limited by extended defects, particularly threading dislocations and vacancy dislocation complexes, which scatter carriers, perturb color-center environments, and degrade quantum coherence [2-5]. Dislocations instill long-range strain fields and introduce additional electronic states in the diamond bandgap [5-7]; beyond electronic and spin effects, vacancy and dislocation defects can concentrate shear, seed crack initiation, and reduce realized mechanical performance at the component level [8-10].

Classical crystal-growth and annealing show that defects on {100} and {111} diamond can reorganize via step-flow, kink motion, and recovery of dislocation networks, providing pathways to reduce energy through defect annihilation or rearrangement [11]. As-processed CVD plates that are mechanically polished (MP) and laser polishing (LP) frequently exhibit high near-surface defectivity, including micro-grooves, micro-cracks, amorphous rims, and residual shear fields that degrade optical and electronic performance and increase graphitization risk at device-relevant depths [12]. Neutral vacancies (GR1) produce optical absorption that anneals near 400 °C, consistent with defect recombination or complexing [13]. Dislocation networks have been proposed to contribute to A-band luminescence and to form extended in-gap states along vacancy-like dislocation cores, yielding an impurity band a few electron volts above the valence band [7,14-16]; suppressing vacancy dislocation states improves room-temperature conductivity and uniformity [3].

From a manufacturing viewpoint, a key challenge is converting commercially available diamond plates into device-ready surfaces over large areas. In this manuscript, industrial-grade diamond denotes plates that meet baseline dimensional and optical specifications but retain a near-surface defect state, whereas functional or technical grade denotes surfaces whose defect-mediated stress proxies and ISO 25178 topography metrics are sufficiently reduced to support repeatable downstream processing. A scalable sub-melt post-process that suppresses near-surface and grain-boundary defect networks is therefore attractive for both high-defect single-crystal diamond (for example heteroepitaxial CVD) and polycrystalline diamond films.

Pulsed-laser processing provides a controllable non-equilibrium route to drive carbon bonding and microstructure by brief heating and rapid quenching. In diamond, nanosecond pulses can generate stress-wave fields and transient bond rehybridization, while higher energy densities can access melt-mediated pathways such as Q-carbon and subsequent diamond formation under appropriate undercooling and pulse conditions at ambient pressure [18-22]. By contrast, sub-melt picosecond to nanosecond excitation can reduce defect densities through stress-assisted bond rotations, vacancy diffusion, and local pressure spikes on short time scales [18,23], and related shear-driven sp2-to-sp3 conversion mechanisms highlight the role of stress in reconfiguring bonding far from equilibrium [24]. Laser-induced evolution in amorphous carbon films further illustrates broader photon-driven reordering pathways [25].

Here we study single-pulse and two-pulse PLA on intrinsic single-crystal diamond with mechanically polished and laser-processed surface damage. We quantify near-surface smoothing and densification by white-light interferometry (ISO 25178 metrics), evaluate strain relaxation and effective depth using depth-resolved Raman mapping, and interrogate local strain structure via cross-sectional STEM and GPA. We develop a minimal model that couples an order parameter for diamond-like versus graphitic-like bonding to intermediate-band carrier populations and elastic stress relief, linking defect-selective photon–phonon excitation to densification in the sub-melt regime [17,18-21,25].

## 2. Thermodynamic and geometric framework (summary)

### 2.1 Summary

We interpret the interferometric and spectroscopic changes using a minimal thermodynamic framework in which the near-surface driving force decreases through a combination of reduced true surface area and reduced defect-mediated strain energy during sub-melt processing. Detailed energetic construction, defect parameterization, and intermediate-band absorption rationale are provided in the Supplementary Information (Section S1). Here we retain only the governing relations needed to map theory to observables.

### 2.2 Governing relations and observable mappings

Surface smoothing is described by a Mullins-type surface diffusion relation (Eq. 9), motivating the use of ISO 25178 gradient and area metrics $Sdq$ and $Sdr$. Here, $Sdq$ is the root mean square gradient of the scale-limited surface (RMS surface slope, a measure of average local steepness), and $Sdr$ is the developed interfacial area ratio, defined as the percent increase in true surface area relative to the projected area. These are reported alongside amplitude parameters $Sa$ and $Sq$.

$$\frac{\partial h}{\partial t} = -B(T)\nabla^4 h, \quad B(T) \propto \frac{D_s(T)\gamma_s \Omega^2}{k_B T}, \tag{9}$$

Here $D_s$ is the surface diffusivity, $\gamma$ is the surface energy, $\Omega$ is the atomic volume, $k_B$ is Boltzmann's constant, and T is the effective temperature during the thermal spike.

To quantify densification from interferometry, we convert ISO 25178-2 functional volumes into an effective thickness change over the scan area A (Eq. 11), enabling nm-scale compaction to be expressed as an absolute displaced volume.

$$\Delta V_{dens} = [(V_{mp} + V_{mc})_{pre} - (V_{mp} + V_{mc})_{post}] + [(V_{vc} + V_{vv})_{pre} - (V_{vc} + V_{vv})_{post}] \quad (\frac{\mu m^3}{\mu m^2}) \tag{11}$$

$$\frac{\Delta V_{dens}}{A} = \int_0^h \Omega_d [c_d^{(0)}(z) - c_d^{(N)}(z)]dz$$

By construction, $\Delta t_{eff} > 0$ indicates a net collapse of peaks and deep valleys and void filling, consistent with densification. This metric is used in Section 4.1 to connect ISO 25178 volume changes to an effective densified thickness.

## 3. Method and Materials

### 3.1 Diamond samples

Two single-crystal CVD diamond coupons (URR Manufacturing, Mumbai, India; nominal 8 mm × 8 mm × 0.7 mm) were used. The as-received surfaces were mechanically polished and laser diced (no CMP), exhibiting dense polishing striations, edge chatter/micro-step features, and residual particulates consistent with MP/LP-induced damage in CVD diamond [12]. Samples were ultrasonically cleaned for 10 min in acetone followed by 10 min in isopropyl alcohol, then dried in $N_2$ air.

Surface topography was measured pre-PLA and post-PLA at identical stage coordinates using a Sensofar optical profilometer in white-light VSI mode (SensoSCAN v6.7) with a DI 10× objective (1.3 μm px$^{-1}$) over a ~1.75 mm × 1.32 mm field of view and a 50 μm Z-scan range. Height maps were leveled by least-squares plane removal; isolated dropouts and dust were masked, with no additional smoothing. ISO 25178 areal parameters (Sa, Sq, Sp, Sv, Sz, Sdq, Sdr) and ISO 25178-2 volume parameters (Vmp, Vmc, Vvc, Vvv; 10 to 80% material-ratio limits) were computed from the leveled maps. For SCD-1, all 30 regions of interest (ROIs) were measured pre and post; percent changes were computed per ROI as ΔX (%) = (X_pre − X_post)/X_pre × 100 and summarized in Fig. S2. Calibration was verified against a step-height standard at the start of each session [30-32].

Raman spectra were acquired pre- and post-PLA at the same ROIs using a Horiba XploRA (532 nm excitation, 80 cm spectrograph, 1800 gr mm$^{-1}$ grating) with a 100× oil-immersion objective (NA = 1.3) and a 25 μm confocal pinhole. Post-PLA depth-resolved line scans were collected and corrected for refractive-index mismatch; depth axes were registered so z = 0 corresponds to the surface. Spectra were baseline-subtracted and the diamond line (~1332 cm$^{-1}$) was fit with a Gaussian to extract peak position, FWHM, and integrated area. Tile 2 (cleaning-only exposure) served as the depth-matched control.

Cross-section lamellae were prepared from SCD-1 at selected ROI such that the lift-outs intercepted polishing striations and traversed the center of the laser spot. TEM foils were prepared on an FEI dual-beam FIB/SEM using a standard ex-situ lift-out protocol. After ROI localization in SEM, the surface was protected with an electron-beam–deposited carbon/Pt cap (~100–150 nm) followed by an ion-beam Pt strap (~1–2 μm). Trenches were milled at 30 kV Ga$^+$ (7–21 nA); lamellae were thinned to ~60–100 nm and final-polished at 5 kV then 2 kV to minimize amorphization and Ga implantation. Imaging was performed on an FEI Tecnai Osiris FEG TEM at 200 kV using BF/HRTEM and STEM (including HAADF where noted).

Geometric phase analysis (GPA) was performed on high-resolution STEM images to quantify local lattice distortion and in-plane strain. Raw HAADF (and where applicable BF) images were analyzed in Fiji (ImageJ) using the Strain++ plugin. An FFT was computed, and one or two non-collinear Bragg reflections were isolated with circular masks, inverse-transformed to obtain Bragg-filtered images/phase maps and referenced to a locally uniform region to remove global phase ramps (tilt/drift). Displacement-gradient fields were differentiated to obtain in-plane strain components, and maps were reported on a fixed strain scale for inter-ROI comparison. Analyses were restricted to high-confidence regions defined by the Bragg-filtered amplitude, excluding low-signal borders.

### 3.2 Laser processing

Pulsed-laser annealing (PLA) was performed at Lawrence Livermore National Laboratory using a frequency-doubled Nd:YAG source (Quanta Ray GCR Pro at 532 nm) delivering 8 ns pulses at 10 Hz to a ~1.100 mm $1/e^2$ beam diameter at the sample plane (Fig. 1), , as measured by a beam profiler at an equivalent plane. As the GR1 center in diamond was documented to have excitation lifetime of approximately 1.1ns at RT [33] with long non-exponential recovery time (tens of milliseconds) of the induced absorption at 532 nm wavelength [13], the experimental parameters were designed to deliver nominal 8 ns pulses at 620 ms inter-shot time, and ~3s inter-site time, such that pulse duration was on the order of excitation times and interpulse cooling time was greater than expected recovery times. Site locations were programmed on 1 mm centers in a grid covering the samples, 7 x 7 for SCD1 and 7 x 5 for SCD-2 (2 rows used for setup). Prior to testing a ratio was measured between the sample plane energy and reference energy reflected off an uncoated fused silica wedge, measurable during testing. The wedge reflects light into a beam profiler positioned at an equivalent plane, such that the energy and beam profile can be measured on every shot during the test. These are combined to produce a fluence map, that is overlaid with sample data using registration marks. Fluence was stepped across the grids from ~17 to 32 J cm$^{-2}$. SCD-1 received one pulse per site, while SCD-2 received two pulses per site.

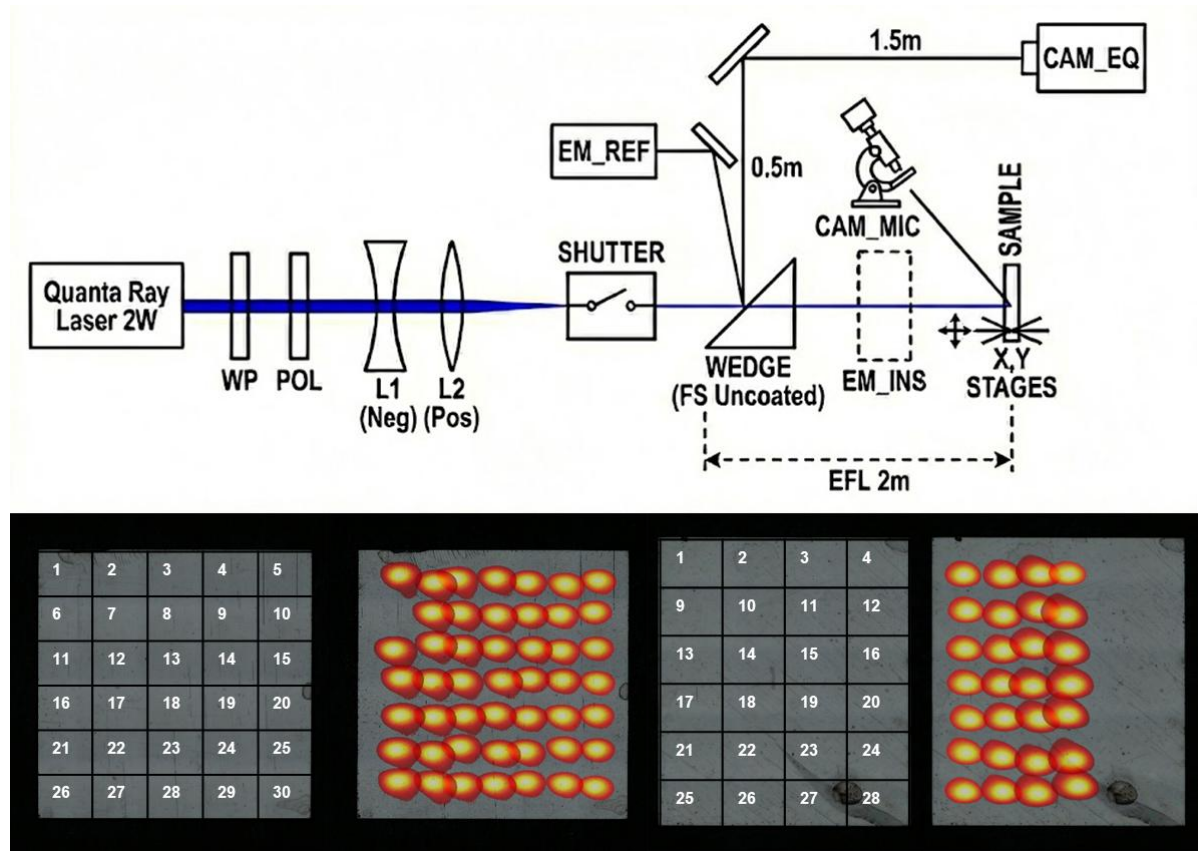

*Figure 1. (Top) Schematic Layout: Quanta Ray Laser (2W), Waveplate (WP), Polarizer (POL), Negative Lens (L1), Positive (L2), Shutter (SHUTTER), Uncoated Fused Silica Wedge (WEDGE), Reference Energy Meter (EM_REF), Camera in Equivalent Plane (CAM_EQ), Insertable Energy Meter (EM_INS), Sample (SAMPLE), X, Y Stages, Camera Microscope System (CAM_MIC), Effective Focal Length (EFL). (Bottom, left to right) Numbered 5×6 grid used to register sites; exposure map for SCD-1 (one pulse per site); grid for SCD-2; exposure map for SCD-2 (two pulses per site). Fluence was stepped from ~17 to 32 J cm⁻².*

## 4. Results and Discussion

We evaluate nanosecond pulsed-laser annealing (ns-PLA) on two single-crystal CVD diamond coupons using three complementary tools: 1) white-light interferometry (ISO-25178 areal and functional volume parameters), 2) depth-resolved Raman of the 1333 cm⁻¹ diamond line (peak position and FWHM), and 3) cross-sectional (S)TEM/GPA at selected sites. Across representative ROIs in the explored sub-melt regime, interferometry shows large reductions in height dispersion and developed area together with a consistent collapse of peak/valley and void volume terms, while Raman shows earlier recovery toward bulk linewidth and reduced near-surface broadening relative to a cleaning-only control.

### 4.1 Effect of PLA on micro and nano-geometries

Figure 2 shows paired VSI height maps (identical stage coordinates) for representative ROIs on SCD 1 before and after single-pulse PLA. Post-PLA maps show a narrower height distribution and reduced ridge-to-valley contrast across the field of view. Table 1 quantifies these changes for ROIs 25, 28, and 29, where Sa and Sq decrease by approximately 48 to 57% together with reductions in Sdq and Sdr. Using Eq. 11 and ISO 25178-2 volumes, the effective densification thickness $\Delta V_{dens}$ is 4.6 nm (ROI 25) and 5.1 nm (ROI 28). Since $\Delta V_{dens}$ is a volume volume-to-surface quantity (µm³/µm²), $\Delta V_{dens} = \Delta t_{eff}$, an effective thickness quantity (µm). Converting this nm-scale thickness signature into absolute volume using $\Delta V_{total} = \Delta t_{eff} \times A$ gives $\Delta V_{total}$ = 10,626 µm³ for ROI 25 and implies A ≈ 2.31 mm², consistent with the interferometry footprint; thus a few nanometers averaged over a millimeter-scale area corresponds to a ~10⁴ µm³-class collapse in void-associated volume, consistent with net compaction of the polishing-damaged layer rather than mere lateral translation of striations. Across the full SCD-1 grid (n = 30), Sa decreases at 29/30 ROIs (median ΔSa ≈ 44%, IQR ≈ 34 to 51%) and Sq decreases at 26/30 ROIs (median ΔSq ≈ 37%, IQR ≈ 24 to 51%) (Fig. S2). Texture metrics also shift in the expected direction, with reduced Sdq at 22/30 ROIs and reduced Sdr at 23/30 ROIs; 10/30 ROIs show drops of 40% or more in Sa, Sq, and Sdq, consistent with a strongly responsive smoothing regime dominated by reduced surface slopes and excess areas that act as geometric stress concentrators.

Table 1 quantifies these changes for SCD-1 (ROIs 25/28/29): Sa and Sq decrease by approximately 48–57%, with corresponding decreases in slope and developed area (Sdq and Sdr). The calculated $\Delta V_{dens}$ for ROIs 25 and 28 was observed to be -4.6 nm and -5.1 nm, respectively, obtained from eq.11 using the ISO 25178-2 volume parameters. Over the scan area, net volume reduction ($\Delta V_{dens} \times A$) was calculated to be 10,626 µm³, supporting a densification/compaction component rather than simple redistribution of roughness. To evaluate repeatability across the full processed grid (rather than only representative panels), we computed ΔSa and ΔSq for all SCD-1 ROIs (n = 30) and plotted the distributions (Fig. S2). Sa shows a reduction at 29/30 ROIs (median ΔSa ≈ 44%, IQR ≈ 34–51%), while Sq shows a reduction at 26/30 ROIs (median ΔSq ≈ 37%, IQR ≈ 24–51%). These distributions demonstrate that the reduction in roughness values reported in Table 1 and illustrated in Fig. 2 are not isolated outcomes, but are broadly reproducible across the grid within this processing window. In addition to amplitude parameters (Sa, Sq), the texture metrics Sdq (RMS surface slope) and Sdr (developed interfacial area ratio) are included because they capture changes in local gradients and true surface area; across SCD-1, 22/30 ROIs exhibit reduced Sdq and 23/30 ROIs exhibit reduced Sdr (Fig. S2), and 10/30 exhibited drops of 40% or more on Sa, Sq, and Sdq, supporting smoothing via reduced surface slope/area rather than simple feature translation.

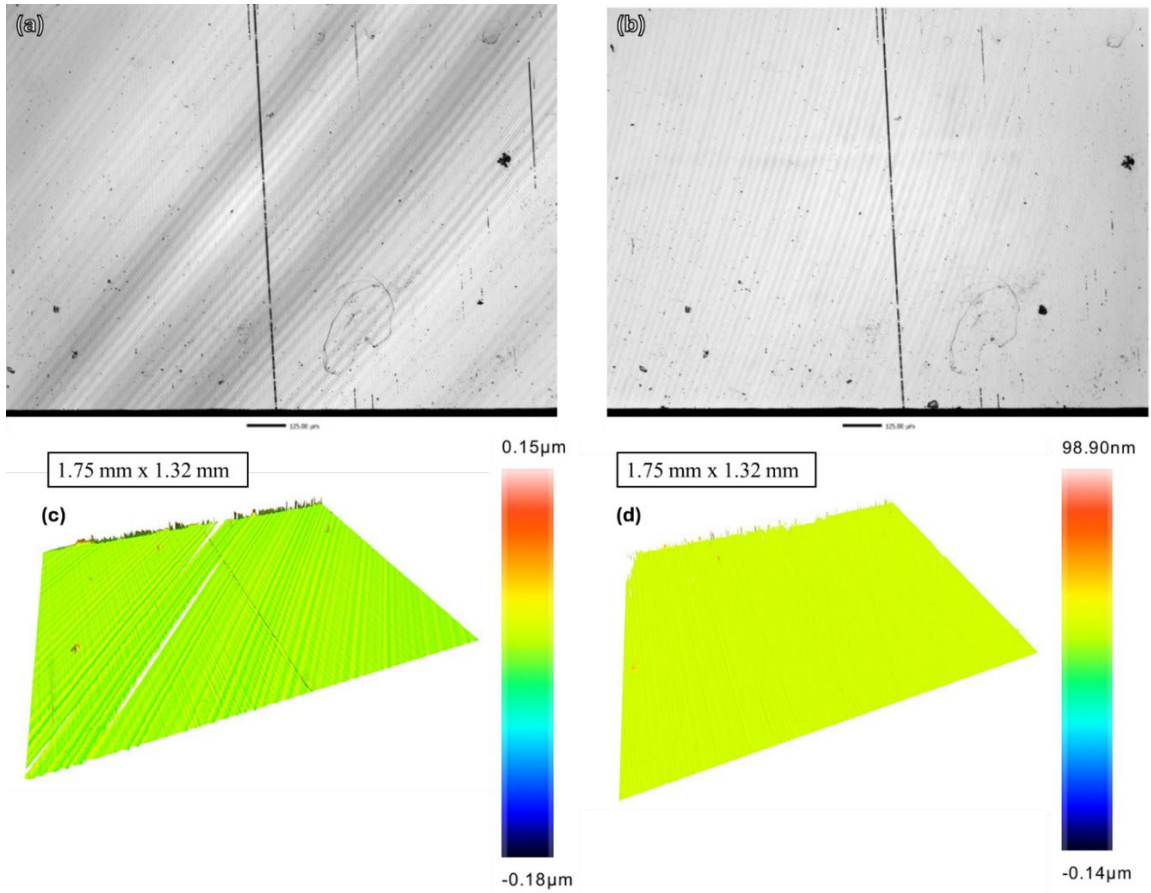

*Figure 2. White-light interferometric height maps (Sensofar S Neox, VSI) of the same ROI before and after single-pulse PLA (SCD-1). Color bars indicate out-of-plane height relative to the local mean (bright = higher, dark = lower).*

Figure 3 shows the analogous comparison for SCD-2 (two pulses per site). The same directional changes are observed (flattened groove field and reduced ridge–valley contrast), with stronger response at higher-fluence sites. Consistent with SCD-1, the ISO-25178 height parameters decrease markedly (Table 1), where the two-pulse case gives higher $\Delta V_{dens}$ than the one-pulse case, (6.5 nm vs 4.6 & 5.1 nm), yet sublinear, consistent with the model's diminishing returns after the first pulse reduces defect-mediated absorption. Smoothing metrics drop strongly, matching the Mullins prediction for a brief high-$T$ window per pulse.

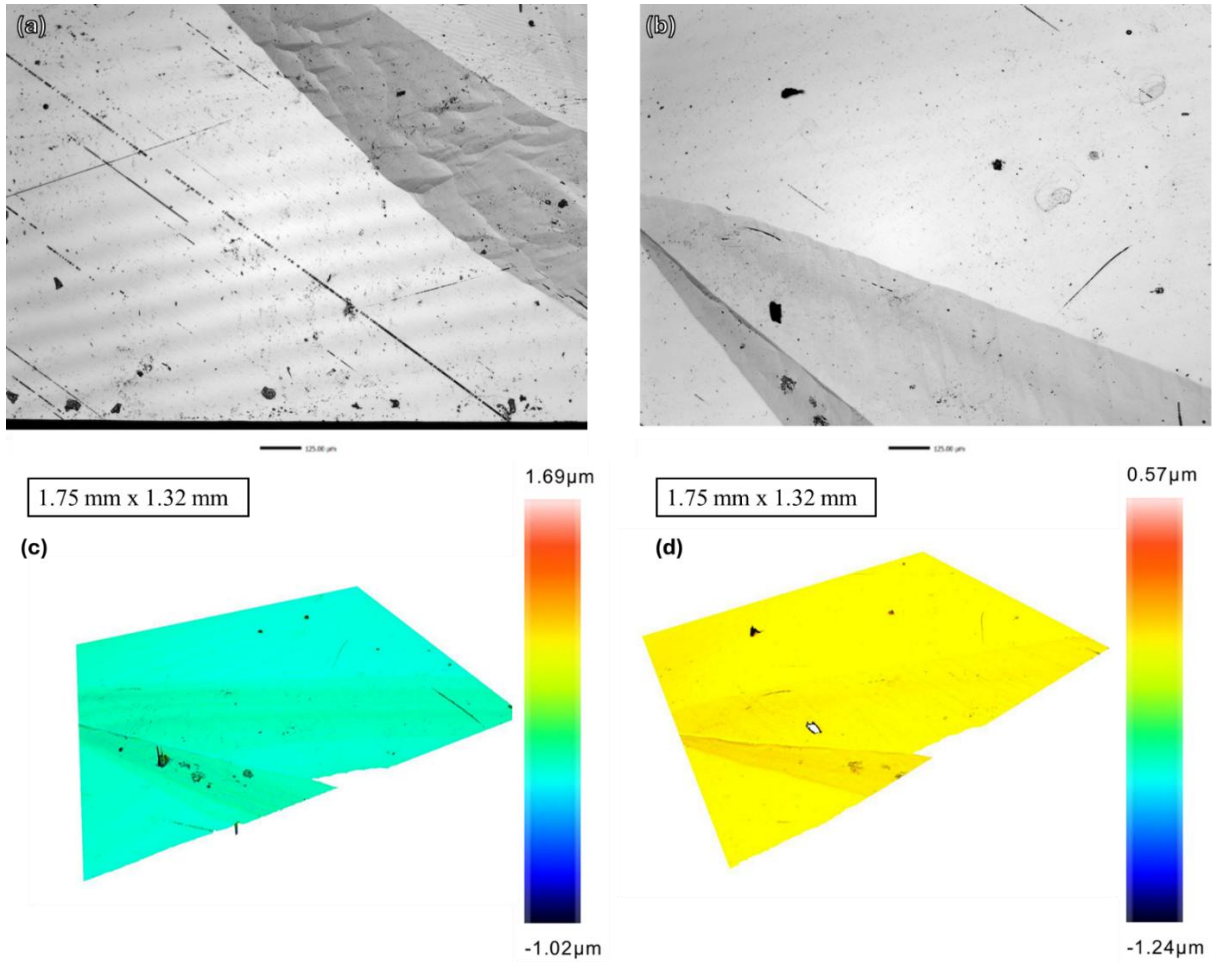

*Figure 3. White-light interferometric height maps (Sensofar S Neox, VSI) of the same ROI on SCD-2 before and after two-pulse PLA. Color bars indicate out-of-plane height relative to the local mean (bright = higher, dark = lower).*

Functional volume parameters provide an integrated measure of peak and valley content. As summarized in Table 1, void-type terms (Vv, Vvc, Vvv) decrease substantially after PLA, and the associated material terms (Vmp, Vmc) decrease in the same direction. The simultaneous collapse of valley- and peak-associated volumes is consistent with rounding/compaction of near-surface topography (and reduced void space) rather than a simple lateral translation of features.

**Table 1**

Summary of the changes to ISO 25178 & ISO 25178-2 parameters for select ROIs on SCD1 & SCD2

| | Sample | SCD 1 | | | SCD2 | | Sample | SCD1 | SCD1 | SCD2 |
|---|---|---|---|---|---|---|---|---|---|---|
| | ROI | Δ 29 (%) | Δ 28 (%) | Δ 25 (%) | Δ 19 (%) | | ROI | Δ 28 (%) | Δ 25 (%) | Δ 19 (%) |
| ISO 25178 Surface Parameters | Sa (nm) | -48.5 | -49.1 | -51.2 | -52.3 | ISO 25178-2 Volume Parameters | Vmp ($\mu m^3/\mu m^2$) | -50.0 | -50.0 | -57.1 |
| | Sq (nm) | -48.3 | -51.5 | -57.1 | -30.4 | | Vmc ($\mu m^3/\mu m^2$) | -46.2 | -47.1 | -41.7 |
| | Sdq (°$\mu m^{-1}$) | -33.3 | -42.3 | -62.7 | -38.4 | | Vvc ($\mu m^3/\mu m^2$) | -48.1 | -47.9 | -61.4 |
| | Sdr (%) | -50 | -66.7 | -88.9 | -63.8 | | Vvv ($\mu m^3/\mu m^2$) | -62.5 | -62.5 | -60.0 |
| | Sp (nm) | -1.3 | -35.6 | -53.6 | -66.1 | | Vv ($\mu m^3/\mu m^2$) | -48.4 | -50.0 | -61.0 |

## 4.2 Effect of PLA on characterized crystal lattice dynamics

Figure 4 compares depth profiles of the diamond Raman line for the cleaning-only control (tile 2) and PLA-processed ROIs (tiles 25, 28, 29). All sites exhibit broader linewidth near the surface that narrows toward bulk values with depth. After PLA, two consistent effects are observed: (i) recovery to FWHM < 2.70 cm$^{-1}$ occurs approximately 0.5 to 1.5 µm earlier than the control, and (ii) at matched depths the PLA ROIs exhibit narrower FWHM by approximately 0.07 to 0.10 cm$^{-1}$. The depth at which the PLA-induced FWHM improvement becomes indistinguishable from the cleaning-only control corresponds to approximately 2 to 3 µm after refractive-index correction, defining an effective processing depth on micron scales. Peak-position profiles show shallower near-surface excursions and quicker return to the bulk value after PLA, consistent with partial relief of near-surface strain and defect-related perturbations. Peak-area roll-off curves overlap after alignment, supporting depth-matched comparisons. No G peak is observed, consistent with the absence of graphitization in this sub-melt window [34-37].

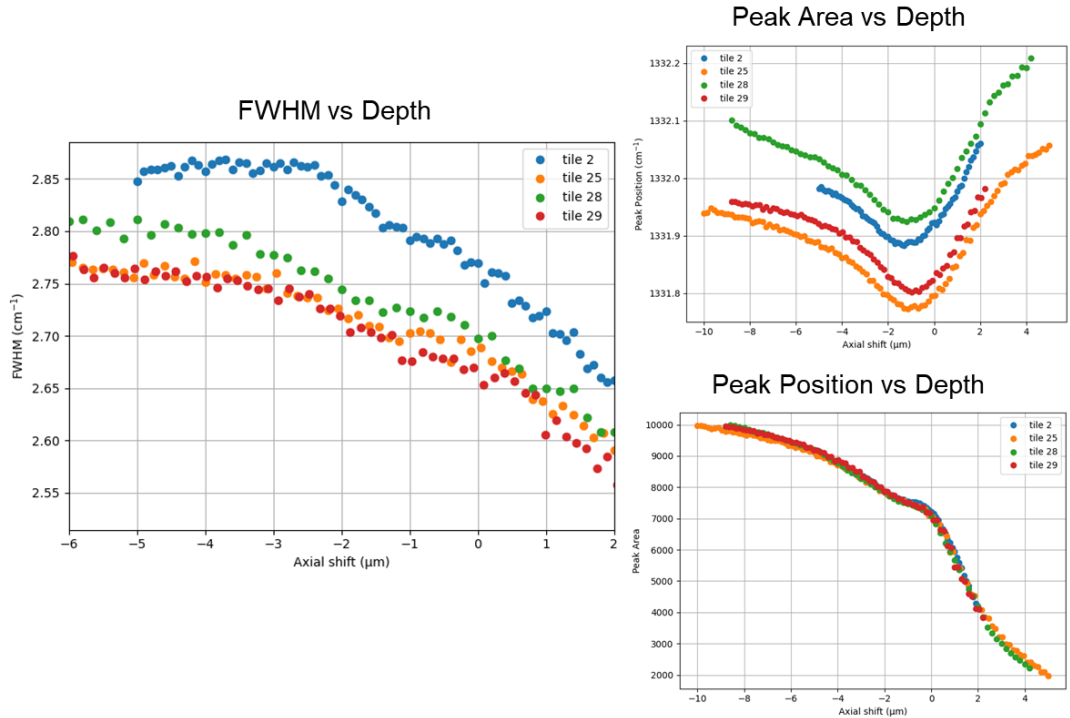

*Figure 4. Depth-resolved Raman of the diamond line acquired with 532 nm excitation (Horiba XploRA; 100× oil-immersion objective; NA 1.3; 25 μm pinhole). Axial depth was corrected for refractive-index mismatch and registered so z = 0 corresponds to the surface.*

To connect these spectroscopic trends to microstructure, we prepared cross-section lamellae from a high-response ROI for (S)TEM and GPA analysis.

4.3     Effect of PLA on characterized crystal structure

Figure 5 shows cross-sectional STEM from a representative processed ROI (tile 28). A thin amorphous rim at the outermost surface is attributed to FIB preparation, while the underlying diamond remains continuous and crystalline over the probed depth. Beneath the cap, contrast variations decay gradually with depth rather than forming a sharp interface, consistent with a shallow gradient (for example strain or density) within the near-surface region rather than a distinct secondary phase or polycrystalline regrowth. No edge dislocations, stacking faults, nor grain boundaries were observed; (such structures can generate stress in the lattice and serve as crack nucleation points that reduce interfacial toughness) [38]. The observed microstructure is consistent with the Raman-inferred effective depth.

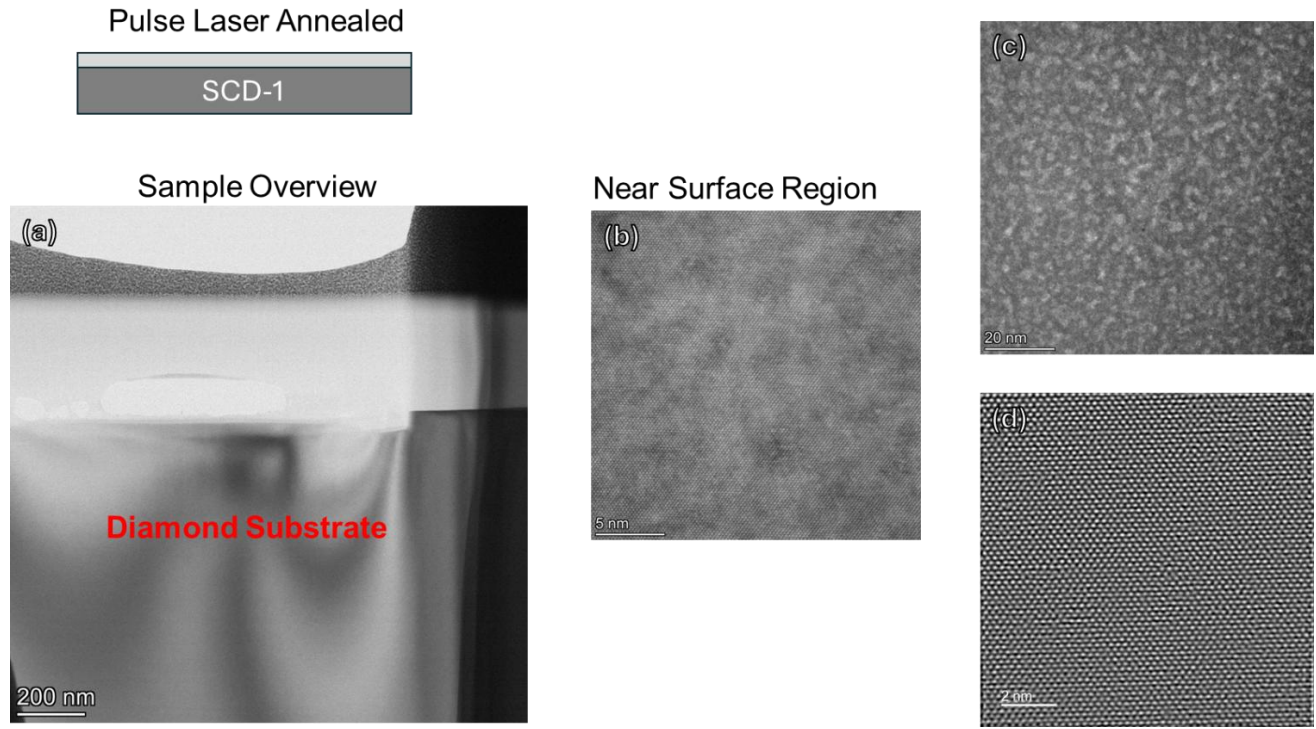

*Figure 5.  Cross-sectional STEM of pulsed-laser-annealed single-crystal diamond (SCD-1, ROI 28). (a) Low-magnification BF-STEM overview of the FIB lift-out acquired from the near-surface region. (b) Higher-magnification BF-STEM near the substrate surface. (c) Intermediate-magnification HAADF-STEM image showing near-surface contrast at 20 nm scale. (d) Atomic-resolution BF-STEM image with FFT-filtered reconstruction highlighting lattice periodicity.*

Geometric phase analysis (GPA) of atomic-resolution STEM images (Fig. 6) shows low-amplitude, smoothly varying strain fields over the analyzed region. Within the current field of view, we do not observe prominent localized dipole-like strain singularities that would be expected from nearby dislocation cores, suggesting that strong near-surface strain concentrators are reduced and/or redistributed after processing. This observation is consistent with the Raman linewidth narrowing and interferometry-calculated compaction.

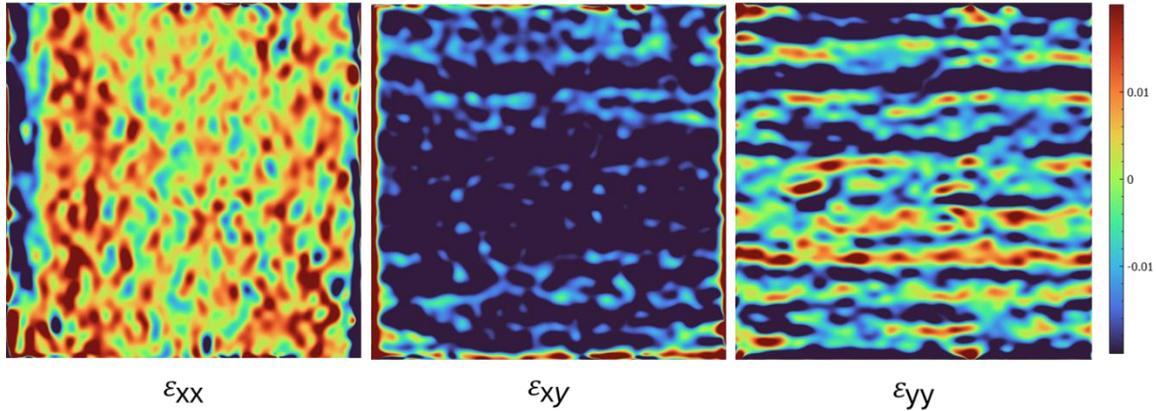

*Figure 6: GPA strain components (εxx, εxy, εyy) computed using Strain++ with a ±0.02 strain display limit (color bar).*

## 5. Discussion and conclusion

Prior studies of laser-induced modification of diamond have focused primarily on high-fluence melting/ablation or ultrafast pulse regimes used to generate color centers [39,40]. At high fluence, nanosecond and excimer pulses melt amorphous or graphitic carbon and rapidly quench it into Q-carbon or new diamond phases, but only after passing through a fully molten state that can also generate cracks, graphitized rims, and new extended defects if not carefully controlled.

In contrast, our results show that ns PLA provides a reproducible pathway to reorganize the near-surface defect layer created during mechanical polishing and laser dicing. Across two single-crystal CVD samples and multiple ROIs, VSI measurements reveal large reductions in Sdq, Sdr, Sp, and Sz, together with a 57 to 60% decrease in areal void volume at fluences above threshold. The effective densification depth is only ~1 nm near threshold and ~4 to 6.5 nm in strongly responsive regions, yet this shallow compaction produces substantial smoothing over millimeter-scale areas by near-surface mass transport that fills monomolecular pits and reduces terrace, kink, and ledge relief. In classical surface-defect models, such surface steps can be coupled to bulk defect structure, including dislocation outcrops and near-surface point-defect populations that generate localized elastic strain. Thus, the observed collapse in Sdq and Sdr is consistent not only with geometric rounding of polishing-induced micro-ledges, but also with reduction or redistribution of subsurface strain concentrators through point-defect migration and partial relaxation of dislocation-associated strain fields.

This interpretation is supported by Raman linewidth recovery extending to approximately 2 to 3 µm and by GPA maps showing smoother, lower-amplitude strain fields after processing, indicating that the response extends well beyond the nm-scale densification signature. This deeper response suggests that surface compaction perturbs the local stress field, enabling relaxation of strain associated with vacancy-dislocation complexes. Cross-sectional STEM shows continuous crystalline diamond without new extended defects, and GPA maps exhibit smoother, low-amplitude strain rather than localized dipole-like strain signatures expected from nearby dislocation cores. Together, these results indicate that ns PLA reduces dislocation-related strain and reorganizes defect-rich networks in the near-surface region, while direct quantification of dislocation density and point-defect concentrations remains an important next step.

From an applications perspective, this behavior is directly relevant to the long-standing challenge of preparing device-ready diamond surfaces at scale. For power and RF electronics, removing near-surface stress concentrators and void-rich layers is expected to increase fracture resistance and reliability under thermal cycling, although direct mechanical testing will be needed to quantify these gains. For quantum and photonic devices based on shallow color centers, near-surface strain and charge noise are known to degrade charge stability and spin coherence, and substantial effort is being spent on surface engineering to mitigate these effects. The ability to reduce dislocation-related strain and smooth topography over several microns without graphitization or aggressive plasma treatments suggests that sub-melt ns-PLA could be inherently scalable to wafer-level treatment and can be embedded in multi-chamber CVD workflows with vacuum transfer for iterative defect mitigation during growth.

Overall, sub-melt ns-PLA offers a practical and scalable post-processing route for improving the functional quality of single-crystal CVD diamond by suppressing near-surface stress concentrators without damaging the lattice. This method has clear potential for preparing device-ready diamond surfaces for electronics, photonics and quantum applications. Future work should quantify mechanical improvements, refine fluence windows for different defect landscapes, and extend the approach to heteroepitaxial and polycrystalline diamond where defect densities are higher.

**Acknowledgments** Work was performed in part in the Stanford Nano Fabrication Facilities (SNF) Labs, which are supported by the National Science Foundation as part of the National Nanotechnology Coordinated Infrastructure under Award ECCS-1542152. This work performed in part under the auspices of the U.S. Department of Energy/National Nuclear Security Administration by Lawrence Livermore National Laboratory under Contract DE-AC52-07NA27344 and Strategic Partnership Program L24336. Raman Microscopy and STEM work was performed by Covalent Metrology.

**Glossary**

| | |
|---|---|
| CVD | Chemical Vapor Deposition |
| MP | Mechanical Polishing |
| LP | Laser Polishing |
| PLA | Pulsed Laser Annealing |
| SCD | Single Crystal Diamond |
| IB | Intermediate Band |
| GB | Grain Boundary |
| G2D | Graphite-to-Diamond |
| FWHM | Full Width at Half Maximum |
| STEM | Scanning Transmission Electron Microscopy |
| BF | Bright Field |
| HAADF | High Angle Annular Dark Field |

**References**

[1] G. Perez, A. Maréchal, G. Chicot, P. Lefranc, P.-O. Jeannin, D. Eon, and N. Rouger, "Diamond semiconductor performances in power electronics applications," Diamond Relat. Mater., vol. 110, (2020)

[2] V. H. Rodgers, L. B. Hughes, M. Xie, P. C. Maurer, S. Kolkowitz, A. C. Bleszynski Jayich, and N. P. de Leon, "Materials challenges for quantum technologies based on color centers in diamond," MRS Bull., vol. 46, no. 7, (2021).

[3] H. Yan, E. Postelnicu, T. Nguyen, S. Corujeira Gallo, A. Stacey, and K. Mukherjee, "Multi-microscopy characterization of threading dislocations in CVD-grown diamond films," Appl. Phys. Lett. 124, 102108 (2024).

[4] D. Li, Q. Wang, X. Lv, L. Li, and G. Zou, "Reduction of dislocation density in single crystal diamond by Ni-assisted selective etching and CVD regrowth," J. Alloys Compd., vol. 960, (2023).

[5] P. Kehayias, M. J. Turner, R. Trubko, J. M. Schloss, C. A. Hart, M. Wesson, D. R. Glenn, and R. L. Walsworth, "Imaging crystal stress in diamond using ensembles of nitrogen-vacancy centers," Phys. Rev. B 100(17), 174103 (2019).
[6] T. Shimaoka, T. Teraji, K. Watanabe, and S. Koizumi, "Characteristic luminescence correlated with leaky diamond Schottky barrier diodes," Phys. Status Solidi A 214(11), 1700180 (2017).
[7] S. Polat Genlik, R. C. Myers, and M. Ghazisaeidi, "Dislocations as natural quantum wires in diamond," Phys. Rev. Mater. 7(2), 024601 (2023).
[8] Qi Zhang et al., Designing Ultrahard Nanostructured Diamond Through Internal Defects and Interface Engineering at Different Length Scales, 170 Carbon 394 (2020), https://doi.org/10.1016/j.carbon.2020.08.036
[9] Anmin Nie et al., Direct Observation of Room-Temperature Dislocation Plasticity in Diamond, 2 Matter 5, 1222–1232 (2020).
[10] Ferdous, S. F., and A. Adnan, "Mode-I fracture toughness prediction of diamond at the nanoscale," *Journal of Nanomechanics and Micromechanics*, 2017, 7(3), 04017017.
[11] Okotrub, A. V., Sedelnikova, O. V., Gorodetskiy, D. V., Gusel'nikov, A. V., Palyanov, Yu. N., & Bulusheva, L. G. (2025). Texture of (100) and (111) faces of annealed diamond crystal. Applied Surface Science, 701, 163270. https://doi.org/10.1016/j.apsusc.2023.163270
[12] Zhang, H., Yan, Z., Zhang, H., & Chen, G. (2025). Graphitization of diamond: manifestations, mechanisms, influencing factors and functional applications. Functional Diamond, 5(1). https://doi.org/10.1080/26941112.2025.2533896
[13] Subedi, S., Fedorov, V., Mirov, S., & Markham, M. (2021). Spectroscopy of GR1 centers in synthetic diamonds. Optical Materials Express, 11(3), 757–767. https://doi.org/10.1364/OME.415334
[14] J.F. Prins. "Increased Band A Cathodoluminescence After Carbon Ion Implantation and Annealing Of Diamond" Diamond Relat, Mater. S (1996) 909.
[15] J. te Nijenhuis, G. Z. Cao, P. C. H. J. Smits, W. J. P. van Enckevort, L. J. Giling, P. F. A. Alkemade, M. Nesládek, and Z. Remeš, "Incorporation of lithium in single crystal diamond: diffusion profiles and optical and electrical properties," Diamond Relat. Mater., vol. 6, (1997)
[16] Liao, Meiyong et al. "Persistent Positive and Transient Absolute Negative Photoconductivity Observed in Diamond Photodetectors." Physical review. B, Condensed matter and materials physics 78.4 (2008).
[17] Khan, A. H., & Kim, T. S. (2025). Advanced co-doping techniques for enhanced charge transport in diamond-based materials. MRS Advances. https://doi.org/10.1557/s43580-025-01206-x
[18] Tian, B., Ma, W., Chen, S., Sun, F., & Wang, X. (2024). Effects of pulsed laser processing on structural evolution of diamonds – A molecular dynamics and experimental study. International Journal of Refractory Metals and Hard Materials, 119, 106560. https://doi.org/10.1016/j.ijrmhm.2024.106560
[19] Khosla, N., Narayan, J., Narayan, R., Sun, X.-G., & Paranthaman, M. P. (2023). Microstructure and defect engineering of graphite anodes by pulsed laser annealing for enhanced performance of lithium-ion batteries. Carbon, 205, 214–225. https://doi.org/10.1016/j.carbon.2023.02.017
[20] Hurtuková, K., Slepičková Kasálková, N., Fajstavr, D., Lapčák, L., Švorčík, V., & Slepička, P. (2023). High-Energy Excimer Annealing of Nanodiamond Layers. *Nanomaterials*, *13*(3), 557. https://doi.org/10.3390/nano13030557
[21] Klaudia Hurtuková et al., Exploring Morphological Diversity of Q-Carbon Structures Through Laser Energy Density Variation, 140 Diamond & Rel. Mat. 110511 (2023), https://doi.org/10.1016/j.diamond.2023.110511
[22] Jagdish Narayan & Anagh Bhaumik (2016) Q-carbon discovery and formation of single-crystal diamond nano- and microneedles and thin films, Materials Research Letters, 4:2, 118-126, DOI: 10.1080/21663831.2015.1126865
[23] Boyu Tian et al., Effects of Pulsed Laser Processing on Structural Evolution of Diamonds—A Molecular Dynamics and Experimental Study, 119 Int'l J. Refract. Metals & Hard Materials 106560 (2024), https://doi.org/10.1016/j.ijrmhm.2024.106560
[24] Dezhou Guo, Kun Luo & Qi An (2024) Shear-promoted graphite-to-diamond phase transition at the grain boundary of nanocrystalline graphite, Functional Diamond, 4:1, 2366807, DOI: 10.1080/26941112.2024.236680
[25] Rivera, A.D.; Hershkovitz, E.; Panoutsopoulos, P.; de Jesus Lopez, M.X.; Simpson, B.; Kim, H.; Narayanan,R.; Johnson, J.; Jones, K.S. Pulsed Laser Annealing of Deposited Amorphous Carbon Films. C2025,11,60. https://doi.org/10.3390/c1103006
[26] W.T. Read and W. Shockley, Phys. Rev. 78, 275 (1950)

[27] Baruffi, C., and C. Brandl. "On the Structure of (111) Twist Grain Boundaries in Diamond: Atomistic Simulations with Tersoff-Type Interatomic Potentials." *Acta Materialia*, vol. 215, 2021, p. 117055.

[28] Wang JB, Yang GW. Phase transformation between diamond and graphite in preparation of diamonds by pulsed-laser induced liquid-solid interface reaction. J Phys Condens Matter. 1999;11(37):7089–7094. doi: 10.1088/0953-8984/11/37/306

[29] Yuhki Tsukada et al., Phase-field simulation of distribution of dislocation density and internal stress in as-quenched martensite in low-carbon steel, 259 Comput. Mater. Sci. 114171 (2025).

[30] Pagani, L., Qi, Q., Jiang, X., and Scott, P. J. (2017). Towards a new definition of areal surface texture parameters on freeform surface. Measurement 109, 281–291.doi: 10.1016/j.measurement.2017.05.028

[31] Czifra Á and Barányi I (2020) Sdq-Sdr Topological Map of Surface Topographies. Front. Mech. Eng. 6:50. doi: 10.3389/fmech.2020.00050

[32] ISO 25178/2-2012 (2012). Geometrical Product Specifications (GPS)–Surface Texture: Areal–Part 2: Terms, Definitions and Surface Texture Parameters

[33] G. Davies, M. F. Thomazs, M. H. Nazares, M. Martin, and D. Shaw, "Vacancy in diamond," 13, 1–6 (1987).

[34] Bhaumik, Anagh, and Jagdish Narayan. "Nano-to-micro diamond formation by nanosecond pulsed laser annealing." Journal of Applied Physics, vol. 126, no. 12, 2019, 125307. AIP Publishing, https://doi.org/10.1063/1.5118890

[35] Karmakar, S., Halim, M. A., Sultana, M., Sarkar, P. K., Emu, I. H., Jaimes-Leal, A. M., & Haque, A. (2024). Enhancing sp³ content in diamond-like carbon thin film electrodes by pulsed laser annealing for durable charge storage performance. Diamond and Related Materials, 146, 111196. https://doi.org/10.1016/j.diamond.2023.111196

[36] Liu, H., Zong, W. & Cheng, X. Load- and Size Effects of the Diamond Friction Coefficient at the Nanoscale. *Tribol Lett* **68**, 120 (2020). https://doi.org/10.1007/s11249-020-01360-3

[37] Lian, Min, et al. *"Enhancing the fracture toughness of polycrystalline diamond by adjusting the transgranular fracture and intergranular fracture modes." International Journal of Refractory Metals and Hard Materials*, vol. 118, 2024, 106490.

[38] Tian, Boyu, et al. "Effects of Pulsed Laser Processing on Structural Evolution of Diamonds – A Molecular Dynamics and Experimental Study." *International Journal of Refractory Metals and Hard Materials*, vol. 119, 2024, p. 106560. Elsevier, https://doi.org/10.1016/j.ijrmhm.2024.106560

[39] J. Smith et al., "Nitrogen-vacancy center formation via local femtosecond laser annealing of diamond," *Carbon*, 246, 120892 (2026).

[40] A. Lee et al., "Single NV centers in diamond produced by multipulse femtosecond laser irradiation," *Diamond and Related Materials*, 155, 112370 (2025).

**Author contributions**

Adam Khan contributed to Funding acquisition (equal), Experimental Conceptualization, Investigation (equal), Supervision (equal), Validation (equal), Visualization (equal), Writing, original draft (equal), and Writing, review and editing (equal). Tae Sung Kim contributed to Investigation (equal), Validation (equal), and Writing, review and editing (equal). Afaq Piracha contributed to Investigation (equal), Validation (equal), and Writing, review and editing (equal). Gabe Guss contributed to Investigation (equal), Ted A. Laurence, Sonny Ly and Thej U. Tumkur contributed to Laser Work Conceptualization, Investigation (equal), Supervision (equal), and Validation (equal).

**Funding** Not applicable: own funds.

**Data availability** The data in this manuscript can be obtained from the corresponding author under reasonable request.

**Declarations** Conflict of interest: The authors do not have any conflict of interest.